# A sub-Mercury-sized exoplanet

Thomas Barclay[1,2], Jason F. Rowe[1,3], Jack J. Lissauer[1], Daniel Huber[1,4], François Fressin[5], Steve B. Howell[1], Stephen T. Bryson[1], William J. Chaplin[6], Jean-Michel Désert[5], Eric D. Lopez[7], Geoffrey W. Marcy[8], Fergal Mullally[1,3], Darin Ragozzine[5,9], Guillermo Torres[5], Elisabeth R. Adams[5], Eric Agol[10], David Barrado[11,12], Sarbani Basu[13], Timothy R. Bedding[14], Lars A. Buchhave[15,16], David Charbonneau[5], Jessie L. Christiansen[1,3], Jørgen Christensen-Dalsgaard[17], David Ciardi[18], William D. Cochran[19], Andrea K. Dupree[5], Yvonne Elsworth[6], Mark Everett[20], Debra A. Fischer[13], Eric B. Ford[9], Jonathan J. Fortney[7], John C. Geary[5], Michael R. Haas[1], Rasmus Handberg[17], Saskia Hekker[6,21], Christopher E. Henze[1], Elliott Horch[22], Andrew W. Howard[23], Roger C. Hunter[1], Howard Isaacson[8], Jon M. Jenkins[1,3], Christoffer Karoff[17], Steven D. Kawaler[24], Hans Kjeldsen[17], Todd C. Klaus[25], David W. Latham[5], Jie Li[1,3], Jorge Lillo-Box[12], Mikkel N. Lund[17], Mia Lundkvist[17], Travis S. Metcalfe[26], Andrea Miglio[6], Robert L. Morris[1,3], Elisa V. Quintana[1,3], Dennis Stello[14], Jeffrey C. Smith[1,3], Martin Still[1,2], & Susan E. Thompson[1,3]

[1] NASA Ames Research Center, Moffett Field, CA 94035, USA
[2] Bay Area Environmental Research Institute, 596 First St West, Sonoma, CA 95476, USA
[3] SETI Institute, 189 Bernardo Ave, Mountain View, CA 94043, USA
[4] NASA Postdoctoral Program Fellow
[5] Harvard-Smithsonian Center for Astrophysics, 60 Garden Street, Cambridge, MA 02138, USA
[6] School of Physics and Astronomy, University of Birmingham, Edgbaston, B15 2TT, UK
[7] Department of Astronomy and Astrophysics, University of California, Santa Cruz, CA 95064, USA
[8] Department of Astronomy, UC Berkeley, Berkeley, CA 94720, USA
[9] Astronomy Department, University of Florida, 211 Bryant Space Sciences Center, Gainesville, FL 32111, USA
[10] Department of Astronomy, Box 351580, University of Washington, Seattle, WA 98195, USA
[11] Calar Alto Observatory, Centro Astronómico Hispano Alemán, C/ Jesús Durbán Remón, E- 04004 Almería, Spain
[12] Depto. Astrofísica, Centro de Astrobiología (INTA-CSIC), ESAC campus, P.O. Box 78, E-28691 Villanueva de la Cañada, Spain
[13] Department and Astronomy, Yale University, New Haven, CT, 06520, USA
[14] Sydney Institute for Astronomy, School of Physics, University of Sydney, Sydney, Australia
[15] Niels Bohr Institute, University of Copenhagen, DK-2100, Copenhagen, Denmark
[16] Centre for Star and Planet Formation, Natural History Museum of Denmark, University of Copenhagen, DK-1350 Copenhagen, Denmark
[17] Stellar Astrophysics Centre, Department of Physics and Astronomy, Aarhus University, Ny Munkegade 120, DK-8000 Aarhus C, Denmark
[18] NASA Exoplanet Science Institute, California Institute of Technology, 770 South Wilson Avenue Pasadena, CA 91125, USA
[19] McDonald Observatory, University of Texas at Austin, Austin, TX, 78712, USA

[20] National Optical Astronomy Observatory, 950 N. Cherry Ave, Tucson, AZ 85719, USA

[21] Astronomical Institute "Anton Pannekoek", University of Amsterdam, The Netherlands

[22] Southern Connecticut State University, New Haven, CT 06515, USA

[23] Institute for Astronomy, University of Hawaii, 2680 Woodlawn Drive, Honolulu HI 96822, USA

[24] Department of Physics and Astronomy, Iowa State University, Ames, IA, 50011, USA

[25] Orbital Sciences Corporation/NASA Ames Research Center, Moffett Field, CA 94035, USA

[26] White Dwarf Research Corporation, Boulder, CO, 80301, USA

**Since the discovery of the first exoplanet[1,2] we have known that other planetary systems can look quite unlike our own[3]. However, until recently we have only been able to probe the upper range of the planet size distribution[4,5]. The high precision of the Kepler space telescope has allowed us to detect planets that are the size of Earth[6] and somewhat smaller[7], but no prior planets have been found that are smaller than those we see in our own Solar System. Here we report the discovery of a planet significantly smaller than Mercury[8]. This tiny planet is the innermost of three planets that orbit the Sun-like host star, which we have designated Kepler-37. Owing to its extremely small size, similar to that of Earth's Moon, and highly irradiated surface, Kepler-37b is probably a rocky planet with no atmosphere or water, similar to Mercury.**

The Kepler spacecraft was launched in 2009 with the goal of determining the frequency of rocky planets in the habitable zone around Sun-like host stars in our Galaxy[9]. Over 150,000 stars are continuously monitored for transits of planetary bodies[10]. Transit-like signals indicative of three planets[11] were detected in the photometric time series data of the star we designate Kepler-37 (also known as KIC 8478994 and KOI-245) over the course of 978 days by the Kepler spacecraft.

Kepler-37 is cooler than the Sun[12]. Stellar spectral templates were fitted to optical spectra of Kepler-37[13,14] to determine initial values of the stellar properties. Despite the low luminosity and low-amplitude oscillations associated with cool main-sequence stars[15], we were able to detect solar-like oscillations in the flux time series of Kepler-37. Kepler-37 is the densest star where solar-like oscillations have been detected and an asteroseismic analysis of these oscillations allowed us to measure precise stellar properties (for more information see Section 1 of the SI). Iterating between the astroseismic and spectroscopically derived stellar properties yielded a final value for the radius of $0.770 \pm 0.026$ $R_\odot$ and mass of $0.802 \pm 0.068$ $M_\odot$. The small uncertainty on the stellar radius is important because the precision with which we can measure the planet radius is limited by the uncertainty on the stellar radius.

We fitted a transiting planet model[17] with three planets to the Kepler data using a Markov-chain Monte Carlo technique[18]. The best fitting model is shown overplotting the transit data in Figure 1. The transit model yields the ratio of planet radius to stellar radius. Using the asteroseismically derived stellar radius, we determined

planet radii $0.303^{+0.053}_{-0.073} R_\oplus$, $0.742^{+0.065}_{-0.083} R_\oplus$ and $1.99^{+0.11}_{-0.14} R_\oplus$ (where $R_\oplus$ is Earth's radius), for the three planets Kepler-37b, c and d, where the quoted uncertainties are the central 68.3% of the posterior distributions (equivalent to 1-σ) and account for both the uncertainty in the ratio of planetary to stellar radius (the dominant term for the inner planet) and the uncertainty on the stellar radius (the dominant term for the outer planet). A full list of planet parameters are given in Table S1 in Section 8 of the SI.

It was not possible to confirm that the three candidate planets are substellar bodies orbiting Kepler-37 using radial velocities[19] or transit timing variations[20]. We therefore explored possible astrophysical scenarios (blends) that can mimic a planet transit across the disc of Kepler-37 using the BLENDER procedure[6,21,22,23]. BLENDER analyses attempt to show that a blend scenario is much less likely than a planet interpretation by creating a wide array of synthetic light curves of various blend scenarios and comparing the goodness-of-fit to the synthetic light curves to that of the true planet model. Inconsistent fits are rejected, as are fits to scenarios that are ruled out by additional information such as high contrast imaging, Kepler Input Catalog[24] colours and high-resolution spectroscopic observations. Details of the supporting data used by BLENDER are provided in Sections 2, 3, 4, and 5 of the SI. The final blend probability is compared with the expected frequency of real planets (known as the planet prior) to calculate the final probability that the candidate is a true planet. Our planet priors are calculated by looking at the frequency of planets with a size within the 3-σ uncertainty range of the measured planet radii. We then apply a correction to the planet prior to account for false positives and incompleteness in the planet catalogue[25].

In this study, we consider blends caused by background eclipsing binaries, background stars transited by a planet and physical companions transited by a planet. For Kepler-37d the only credible blend scenario is a planet transiting a background G or K-type main sequence star with a brightness within 3 magnitudes of Kepler-37, with a similar absolute radial velocity, but not seen by high contrast imaging. The probability of such a source existing is vanishingly small with an upper limit on the probability of $10^{-10}$. We used the Kepler planet catalogue[11] to estimate the occurrence of planets with a similar radius to Kepler-37d, and hence calculate the planet prior. We find a scenario whereby the planet orbits the target star is favoured by a factor of $>10^6$ over false positive scenarios.

The validation of Kepler-37c proved slightly more challenging owing to the difficulty of calculating a planet prior for such a small body. The number of Kepler planet candidates orbiting F-M-type dwarfs with a similar size to Kepler-37c is 46, and we predict that the expected number of false positives in this size range[25] to be 7.36. There are a total of 138,254 dwarf stars observed continuously for the duration of the mission included in the Kepler planet catalogue and therefore the planet prior is (46 - 7.36) / 138,254=2.79 x$10^{-4}$ not accounting for incompleteness. If we consider that a transit signal this small could only be detected transiting 6.5% of targets owing to the noise level in most stars exceeding the signal we detect here, we can boost the planet prior by a factor of 1/0.065=15.4, giving a final planet prior of 4.3 x $10^{-3}$. The blend frequency for this source is calculated by summing up all the individual blend probabilities; planets transiting background stars (1.49 x $10^{-5}$), background eclipsing binaries (1.4 x $10^{-8}$) and planets transiting a physically bound companion (3.3 x $10^{-7}$). Therefore the total blend probability is 1.52 x $10^{-5}$ and the odds ratio in favour of the true planet scenario is 287. As is common in BLENDER analyses, we can increase the probability of the true planet scenario because mutli-planet systems are likely to be coplanar[26,27,28]. This increases the odds ratio by a factor of 8.3 to 2350, equal to a confidence level of 99.95%.

For Kepler-37b, using the planet candidate list[11] to derive a prior is not appropriate because the list is woefully incomplete at such small planet radii. In this work we assume that planets of comparable size to Kepler-37b have the same occurrence rate as Earth-size planets[25]. The size range we use to calculate our planet prior is the 3-σ uncertainty range of the radius of Kepler-37b. This assumption is somewhat conservative as for larger planet sizes the occurrence rate increases with decreasing planet size[4]. There are 167 Earth-sized candidates in the Kepler planet candidate list. However, we expect about 21 of these to be false positives[25]. Given only 15.5% of the stars observed by Kepler are capable of hosting a detectable Earth-sized planet, we find that (167 – 21)/(138,254 x 0.155) = 0.68% of F-K-type dwarf stars are expected to host a *transiting* Earth-sized planet. We therefore take 0.68% to be our planet prior for Kepler-37b. The blend frequency in this analysis is dominated by background eclipsing binaries between 9 and 11.5 magnitudes fainter than the target (the background eclipsing binary blend frequency is 1.9 x $10^{-5}$, while the blend frequency from all other sources is 7.6 x $10^{-6}$). This results in a total frequency of blends of 2.6 x $10^{-5}$. The ratio of the planet prior to the blend frequency yields an odds ratio in favour of the planet interpretation of 262. If we consider the coplanarity of the system, the odds ratio is boosted by a factor of 8.97 to 2350 (a 99.95% confidence in the true planet interpretation).

We ran a dynamical simulation in order to determine whether this system of three planets is stable. For the initial conditions we used the observed orbital period and the epoch of first transit for each of the three confirmed planets and assume coplanar orbits. We adopt masses based on the relation
$M_p/M_\oplus \approx (R_p/R_\oplus)^{2.05}$ [27]. For each planet, we performed 1,000 integrations of 100,000 years each using initially circular orbits for the other planets. All planets were found to be stable over the entire simulation.

While sub-Mercury-sized planets are expected from theory[29] and their space-based detection has previously been predicted[30], our detection of Kepler-37b is quite remarkable given this transit signal would be detected in the data of fewer than 0.5% of stars observed by Kepler. Although the detection of one planet cannot be used to determine of occurrence rates, it does lend weight to the belief that planet occurrence increases exponentially with decreasing planet size.

**References**
1. Mayor, M. & Queloz, D. A Jupiter-mass companion to a solar-type star. *Nature* **378**, 355-359 (1995)
2. Marcy, G. & Butler R. P. A Planetary Companion to 70 Virginis. *Astrophys. J.* **464**, L147-L151 (1996)
3. Fabrycky, D. C. *et al. Architecture of Kepler's* Multi-transiting Systems: II. New investigations with twice as many candidates. *Astrophys. J.* submitted, arXiv: 1202.6328 (2012)
4. Howard A. W. *et al.* The Occurrence and Mass Distribution of Close-in Super-Earths, Neptunes, and Jupiters. *Science* **330**, 653-655 (2010)
5. Howard A. W. *et al.* Planet Occurrence within 0.25 AU of Solar-type Stars from Kepler. *Astrophys. J. Suppl.* **201**, 15 (2012)
6. Fressin F. *et al.* Two Earth-sized planets orbiting Kepler-20. *Nature*, **482**, 195–198 (2012)
7. Muirhead, P. S. *et al.* Characterizing the Cool KOIs. III. KOI 961: A Small Star with Large Proper Motion and Three Small Planets. *Astrophys. J.* **747**, 144 (2012)
8. Archinal, B. A. *et al.* Report of the IAU Working Group on Cartographic Coordinates and Rotational Elements: 2009. *Celestial Mechanics and Dynamical Astronomy* **109**, 101–135 (2011)
9. Borucki, W.J. *et al.* Kepler Planet-Detection Mission: Introduction and First Results. *Science* **327**, 977–980 (2010)
10. Koch, D. G., *et al.* Kepler Mission Design, Realized Photometric Performance, and Early Science. *Astrophys. J.* **713**, L79-L86 (2010)
11. Batalha, N. M. *et al.* Planetary Candidates Observed by Kepler, III: Analysis of the First 16 Months of Data. *Astrophys. J. Suppl.* submitted, arXiv:1202.5852 (2012)
12. Ammons S. M. *et al.* The N2K Consortium. IV. New Temperatures and Metallicities for More than 100,000 FGK Dwarfs. *Astrophys. J.* **638**, 1004–1017 (2006)
13. Valenti, J. A. & Fischer, D. A. Spectroscopic Properties of Cool Stars (SPOCS). I. 1040 F, G, and K Dwarfs from Keck, Lick, and AAT Planet Search Programs. *Astrophys. J. Suppl.* **159**, 141-166 (2005)
14. Buchhave, L. A. *et al.* An abundance of small exoplanets around stars with a wide range of metallicities. *Nature* **486**, 375-377 (2012)
15. Kjeldsen, H. & Bedding, T. R. Amplitudes of stellar oscillations: the implications for asteroseismology. *Astron. Astrophys.* **293**, 87–106 (1995)
16. Ulrich, R. K. Determination of stellar ages from asteroseismology. *Astrophys. J.* **306**, L37– L40 (1986)
17. Mandel, K. & Agol, E. Analytic Light Curves for Planetary Transit Searches. *Astrophys. J.* **580**, L171-L175 (2002)
18. Ford, E. B. Quantifying the Uncertainty in the Orbits of Extrasolar Planets. *Astron. J.* **129**, 1706–1717 (2005)
19. Struve, O. Proposal for a project of high-precision stellar radial velocity work. *The Observatory* **72**, 199-200 (1952)

20. Holman, M. J. & Murray, N. W. The use of Transit Timing to Detect Terrestrial-Mass Extrasolar Planets. *Science 307*, 1288-1291 (2005)
21. Torres, G., Konacki, M., Sasselov, D. D. & Jha, S. Testing blend scenarios for extrasolar transiting planet candidates. I. OGLE-TR-33: a false positive. *Astrophys. J.* **614**, 979-989 (2004)
22. Torres, G. *et al.* Modeling Kepler transit light curves as false positives: rejection of blend scenarios for Kepler-9, and validation of Kepler-9 d, a superearth-size planet in a multiple system. *Astrophys. J.* **727**, 24 (2011)
23. Fressin, F. *et al.* Kepler-10c, a 2.2-Earth radius transiting planet in a multiple system. *Astrophys. J. Suppl.* **197**, 5 (2011)
24. Brown, T. M., Latham, D. W., Everett, M. E. & Esquerdo, G. A. Kepler Input Catalog: Photometric Calibration and Stellar Classification. *Astron. J.* **142**, 112 (2011)
25. Fressin, F. *et al.* The False Positive Rate of Kepler and the Occurrence of Planets. *Astrophys. J.* submitted (2012)
26. Cochran, W. D. *et al.* Kepler-18b, c, and d: A System of Three Planets Confirmed by Transit Timing Variations, Light Curve Validation, Warm-Spitzer Photometry, and Radial Velocity Measurements. *Astrophys. J.* **197**, 7 (2011)
27. Lissauer, J. J. *et al.* Architecture and Dynamics of Kepler's Candidate Multiple Transiting Planet Systems. *Astrophys. J. Suppl.* **197**, 8 (2011)
28. Lissauer J. J. *et al.* Almost All of Kepler's Multiple Planet Candidates are Planets. *Astrophys. J.* **750**, 112 (2012)
29. Chambers, J. Terrestrial Planet Formation. in *Exoplanets* ed. Seager, S. 297-317 (University of Arizona Press, Tucson, AZ; 2011)
30. Borucki, W. J. *et al* Kepler Mission: A Mission to Find Earth-size Planets in the Habitable Zone. In Earths: DARWIN/TPF and the Search for Extrasolar Terrestrial Planets **ESA SP-539**, 69-81 (ESA Publications Division, Noordwijk; 2003)

**Acknowledgements** Kepler was competitively selected as the tenth Discovery mission. Funding for this mission is provided by NASA's Science Mission Directorate. Some of this work is based on observations made with the Spitzer Space Telescope, which is operated by the Jet Propulsion Laboratory, California Institute of Technology under a contract with NASA. Support for this work was provided by NASA through an award issued by JPL/Caltech.
Kepler flux time series data presented in this paper are available from the Mikulski Archive for Space Telescopes (MAST) at the Space Telescope Science Institute (STScI). Funding for the Stellar Astrophysics Centre is provided by The Danish National Research Foundation. The research is supported by the ASTERISK project funded by the European Research Council. E.A. acknowledges support through an NSF Career. DH is supported by an appointment to the NASA Postdoctoral Program at Ames Research Center.

**Author Information** Reprints and permissions information is available at www.nature.com/reprints. The authors declare no competing financial interests. Readers are welcome to comment on the online version of this article at www.nature.com/nature. Correspondence and requests for materials should be addressed to T.B. (thomas.barclay@nasa.gov) or J.F.R (jason.rowe@nasa.gov).

**Author Contributions** T.B. led the work, performed the MCMC transit modelling and wrote the manuscript. J.F.R. discovered Kepler-37b and c and performed the initial analysis of the light curve. J.J.L. provided guidance on the false positive probability, calculated the multiplicity boost and contributed to the manuscript. D.H. discovered the solar-like oscillations and led the asteroseismic analysis. F.F. led the BLENDER analysis. S.T.B. performed pixel level centroid analysis. W.J.C. led the asteroseismic modelling effort. J-M.D. and D.Ch obtained and analysed the Spitzer observations and wrote the section in the SI based on these data. E.D.L. calculated the planetary composition constraints. G.W.M. obtained HIRES spectra and performed the cross-correlation function calculation. F.M. developed a model for assessing false positive probabilities based on colors. D.R. calculated the coplanarity boost used in BLENDER. G.T. developed the BLENDER technique and was involved in the corresponding analysis. El.A. and A.D. obtained and analysed the AO image from the MMT. Er.A. identified additional transiting planet candidates in the Kepler-37 system. D.B. and J.L-B. obtained and analysed the lucky imaging data. S.B., T.R.B., J.C-D, Y.E., R.H., Sa.H., C.K., S.D.K., H.K., Mik.L., Mia.L., T.S.M., A.M., and D.S. were involved in the asteroseismic analysis and modelling. L.A.B. analysed the TrES and HIRES spectra using SPC. J.L.C, M.R.H., J.M.J., T.C.K., J.L., R.L.M, E.V.Q. , J.C.S., M.S. and S.E.T. were involved in the target management, processing, analysis and dissemination of Kepler data. D.Ci., M.E., E.H. and S.B.H. observed and analysed the high contrast imaging data. D.A.F. analysed HIRES spectra using SME. J.C.G developed the Kepler spacecraft photometer electronics, build Keplercam for the KIC and followup spectral observations and developed of the TRES echelle spectrograph at SAO for follow-up observations. E.B.F. contributed to analysis of transit times and eccentricities. J.J.F. assisted in modelling the structure and evolution of the planets. C.E.H. assisted in running BLENDER on the NASA Pleiades supercomputer. A.W.H. and H.I obtained and analysed the HIRES RV data. R.H. manages the Kepler project. D.W.L. obtained TrES spectra.

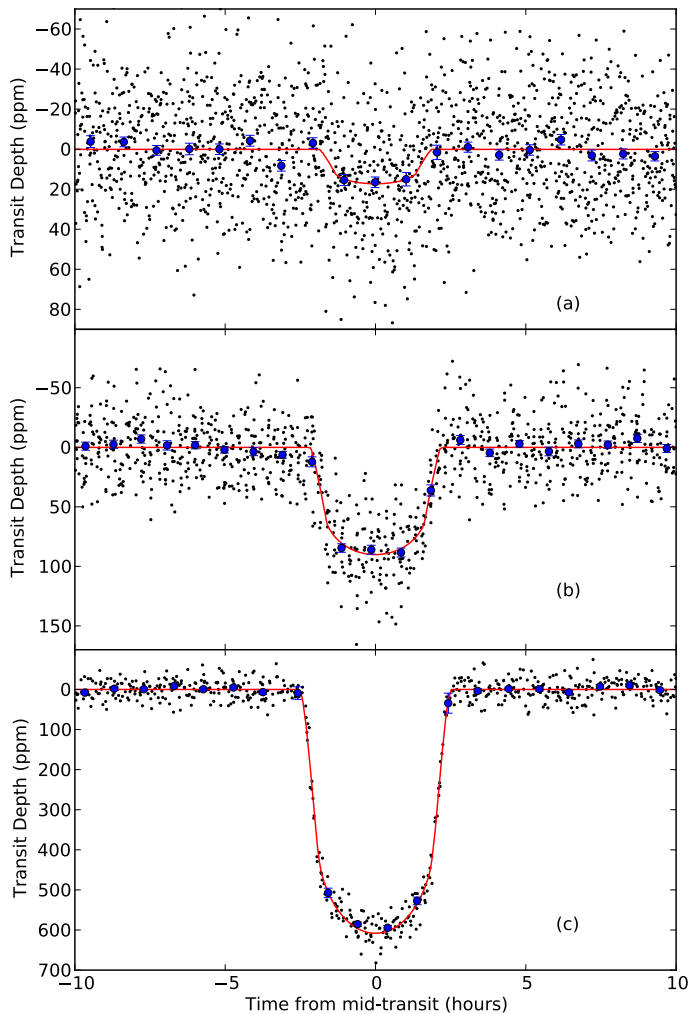

**Figure 1 The transit light curves for the planets orbiting Kepler-37** The transit model fit to the light curve of Kepler-37 was calculated using a four-parameter non-linear limb-darkening model that allowed for eccentric orbits. Limb-darkening coefficients were interpolated in stellar temperature, surface gravity and metallicity to determine the appropriate values for Kepler-37, and these were kept fixed. A Markov-chain Monte Carlo technique was used to sample the planet parameters in order to account for correlated variables. The mean stellar density determined by our asteroseismic analysis was used as a prior in the analysis. The three panels show the transits of planets Kepler-37b (a), Kepler-37c (b) and Kepler-37d (c). The photometric light curve has been folded on the orbital period of the planets. Individual data points are shown as black dots. The blue dots show the data binned with 90, 50 and 30 individual data points making up each binned point for planets b, c and d, respectively. The error bar size is the standard deviation of the data making up that bin divided by $N^{1/2}$, where N is the number of data point in the bin. The best fitting transit model from the Markov-chain Monte Carlo analysis is shown as the red line. The signal to noise of the transits of planets b, c and d is 13, 49 and 282, respectively.

# Supporting Information
# Contents

1. **Determination of the physical parameters of Kepler-37**
2. **Flux centroid analysis**
3. **High resolution spectroscopic observations**
4. **High contrast imaging**
5. **Transits of Kepler-37d at 4.5 microns with Warm-Spitzer**
6. **Constraints on the composition of Kepler-37b from mass loss**
7. **Other planet candidates in the photometric data of Kepler-37**
8. **Stellar and planetary parameters for the Kepler-37 system**

## 1. Determination of the physical parameters of Kepler-37

We detected solar-like oscillations in the short-cadence (1 minute time resolution) flux time series of Kepler-37. In Figure S1 we show a frequency-power spectrum of the Kepler-37 flux time series with the transits having been removed. The spectrum shows a near-regular pattern of overtones that are the signatures of small-amplitude, solar-like oscillations[31]. The dominant frequency spacing is the large separation, $\Delta\nu$, between consecutive overtones of oscillations that have the same spherical (angular) degree, $l$. The spectrum shows several radial ($l$=0) and dipole ($l$=1) modes, which alternate in frequency. The average value of the large frequency separation, $\langle\Delta\nu\rangle$, depends on the mean stellar density[16]. We measured $\langle\Delta\nu\rangle$ [32] for Kepler-37 to be 178.7 ± 1.4 µHz. We also measure an average value for the small frequency separation, $\langle\delta\nu_{01}\rangle$, of 4.08 ± 0.17 µHz, where the small frequency separation is the amount in frequency by which dipole modes are offset from the midpoint of the adjacent radial modes. For main-sequence dwarfs, $\langle\delta\nu_{01}\rangle$ depends largely on the sound speed gradient in the central regions of the star.

Kepler-37 has the largest $\langle\Delta\nu\rangle$ ever measured in a star. Even without further modelling, the observed $\langle\Delta\nu\rangle$ (and also the high frequencies at which the star oscillates, shown in Figure S1) indicates that Kepler-37 is smaller and less massive than the Sun. The determination stellar parameters of from asteroseismology is dependent on stellar evolution models and these require knowledge of the stellar metallicity ( [m/H] ) and effective temperature ($T_{eff}$).

We determined $T_{eff}$ and [m/H] by obtaining spectroscopic observations of Kepler-37 using the HIRES spectrograph[33] mounted on the Keck I telescope on Mauna Kea, Hawaii as well as the fiber-fed Tillinghast Reflector Échelle Spectrograph (TRES) on the 1.5 m Tillinghast Reflector at the Fred Lawrence Whipple Observatory on Mt. Hopkins, Arizona. One iodine-free HIRES template spectrum with a signal-to-noise per resolution element (SNRe) of 257 was acquired on 2 August 2010 and two TRES spectra with an SNRe of 75 and 52, respectively, were acquired on 25 March and 6 April 2010.

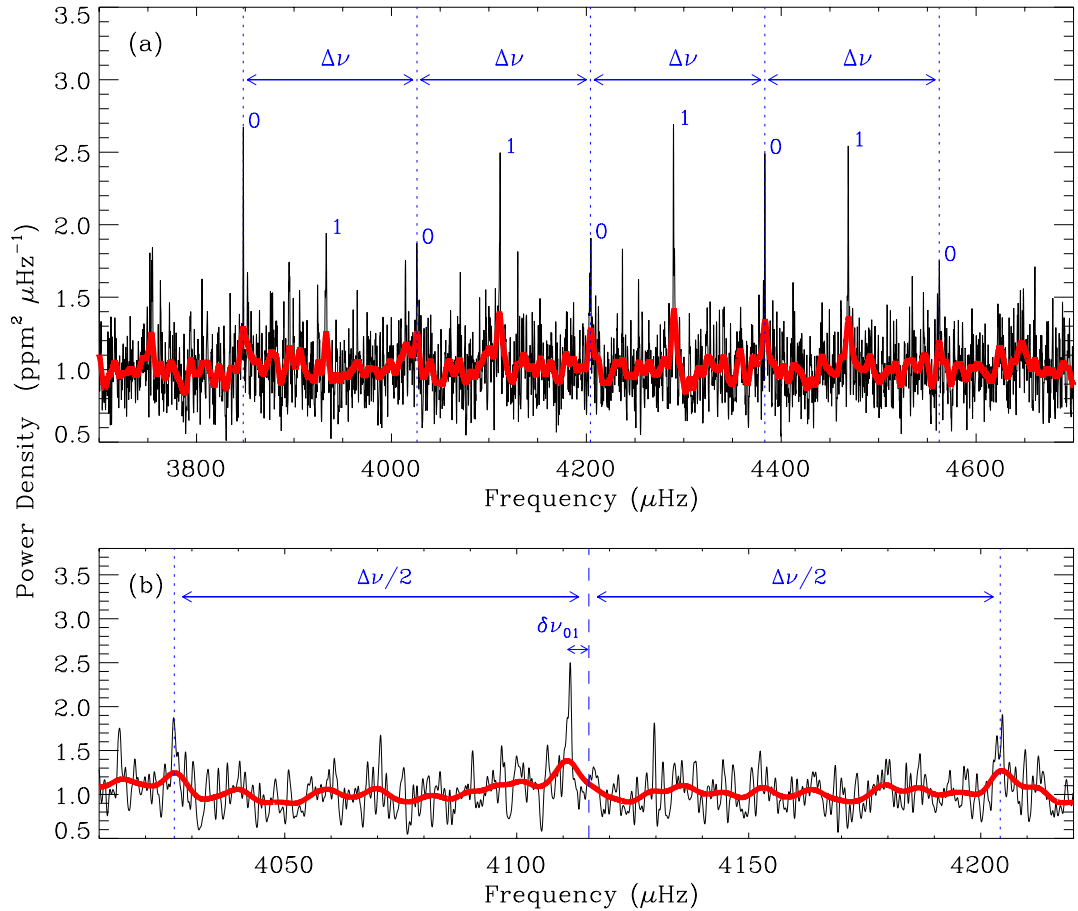

**Figure S1 Oscillation spectrum of Kepler-37.** Panel (a): The power density spectrum after being smoothed by a 0.5-µHz wide boxcar filter (black) and a 5-µHz wide boxcar filter (red). This is based upon 15 months of short cadence (1-minute sampling) Kepler data, collected between March 2010 and June 2011. The annotation marks detected radial modes (having angular degree l=0) and dipole modes (l=1), with large frequency separations, Δν, also shown. Panel (b): Zoom showing an l=1 mode bracketed by the neighbouring l=0 modes, and the small frequency separation $δν_{01}$.

The HIRES spectrum was analysed using SME[13] and all three spectra were analysed using SPC[14]. The SPC parameters used were the weighted average of the parameters from the individual spectra. Initial SME and SPC derived parameters were calculated independently from the asteroseismology. We then used a `grid-based' approach[34,35,36] to formally estimate the stellar properties by searching amongst grids of stellar evolutionary models to get parameters that best fit the two observed seismic frequency separations (large and small). We searched nine different grids of models to account for the effect that systematic differences between the various models had on the inferred stellar properties. The asteroseismic analysis allowed us to fix the surface gravity to log g = 4.5667. Re-running SME and SPC with this constraint yielded a revised $T_{eff}$ and [m/H].

We derive the final stellar parameters by simply taking the average of the SME and SPC values. To account for the differences between the SME and SPC methods we included an uncertainty[37] in $T_{eff}$ of 59 K and in [m/H] of 0.062 dex. These uncertainties are in addition to the formal uncertainties calculated by both techniques and were added in quadrature. Our final $T_{eff}$ = 5417 ± 75 K and [m/H] = -0.32 ± 0.07. Using these final parameters from the spectra we calculated a stellar mass and radius from each of our nine grids. Our final value was calculated by taking the median value of the mass and radius from the various grids while the uncertainty was obtained by adding the standard deviation of the grid masses and radii in quadrature with the formal uncertainties. Our final determination of the stellar radius is 0.772±0.026 $R_\odot$ and the stellar mass is 0.803±0.068 $M_\odot$. We note that the fractional uncertainty on mass is around three times larger than that on the radius. This comes from propagating the uncertainty on mean stellar density to an uncertainty on the modelled mass and radius.

A detailed analysis of the individual mode frequencies will be required to place tight constraints on the age, with preliminary results suggesting an age of around 6 Gyr.

## 2. Flux centroid analysis

For many planets and planet candidates found using Kepler, the spatial location of the flux centre in transit relative to that out of transit provides robust limits on the spatial distance between the source of the transits and the target star[38]. A value not consistent with a zero offset indicates that source of the transits is unlikely to be the target star and hence we have detected a false positive. However, in saturated targets such as Kepler-37 this proves much more challenging. Similarly to the validation of Kepler-21b[49], we are able to use the unsaturated wings of the pixel response function of Kepler-37 to put limits on the separation. We were able to place an upper limit on the separation between the Kepler-37 and a blend mimicking a transit signal of around one pixel for all three candidate planets in the Kepler-37 system. In order to provide a conservative limit we choose 8 arcsec (two pixels) as our upper limit on the offset between Kepler-37 and the source of the transit signals.

## 3. High resolution spectroscopic observations

Relative radial velocities derived from high resolution spectra obtained using the HIRES instrument on the Keck I telescope[32] and the resulting Lomb-Scargle periodogram[40,41] of these data are shown in Figure S2. The periodogram shows no evidence for radial velocity motion with amplitude greater than 7 m s$^{-1}$ on periods up to 500 days. Taking a stellar jitter of 2.0 m s$^{-1}$ [42] and adding that in quadrature to the mean error on our radial velocities, we obtain a mean radial velocity error including stellar jitter of 2.4 m s$^{-1}$. This is consistent with the standard deviation of radial velocities of 3.6 m s$^{-1}$ at the level of 2-σ, indicating there is no measurable radial velocity motion. Additionally, no significant power at any period related to the orbit of any of the three planets is seen.

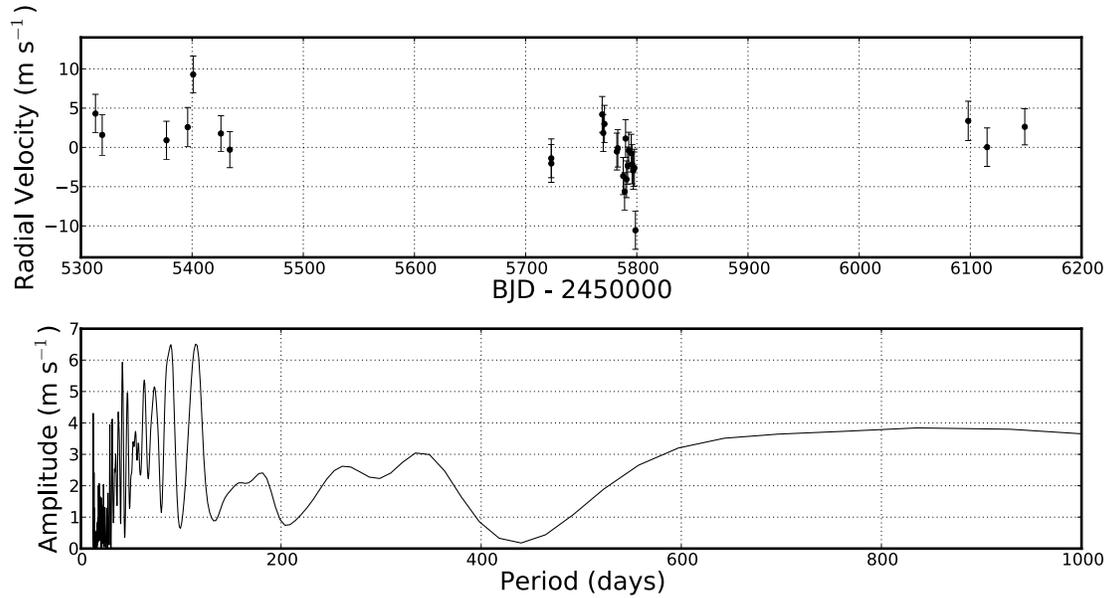

**Figure S2 Radial velocity observations of Kepler-37.** The relative radial velocities derived using the HIRES instrument on the Keck Telescope are shown in the top panel. The observational baseline covers 836 days and uncertainties include both instrumental and stellar jitter. The lower panel shows a Lomb-Scargle periodogram of the radial velocities. There is no evidence for any radial velocity above 7 m s$^{-1}$ over the duration of the observations.

We cross-correlated the HIRES spectra with a template of the Sun's spectrum (using Ganymede as the proxy source). The resultant cross-correlation function (CCF) is shown in Figure S3. The CCF is a single peak, with a width consistent with the widths of the spectral lines of Kepler-37 (and of the Sun). The brightness limits on any companion star can be determined from the smallest detectable second peak in the CCF. No secondary peak is seen in the CCF, down to limits of a 2% of the brightness of the main star, i.e. within 4 magnitudes. We can therefore rule out companions having optical brightness within 4 magnitudes of Kepler-37, except for those stars with very similar absolute radial velocity motion (±10 km s$^{-1}$).

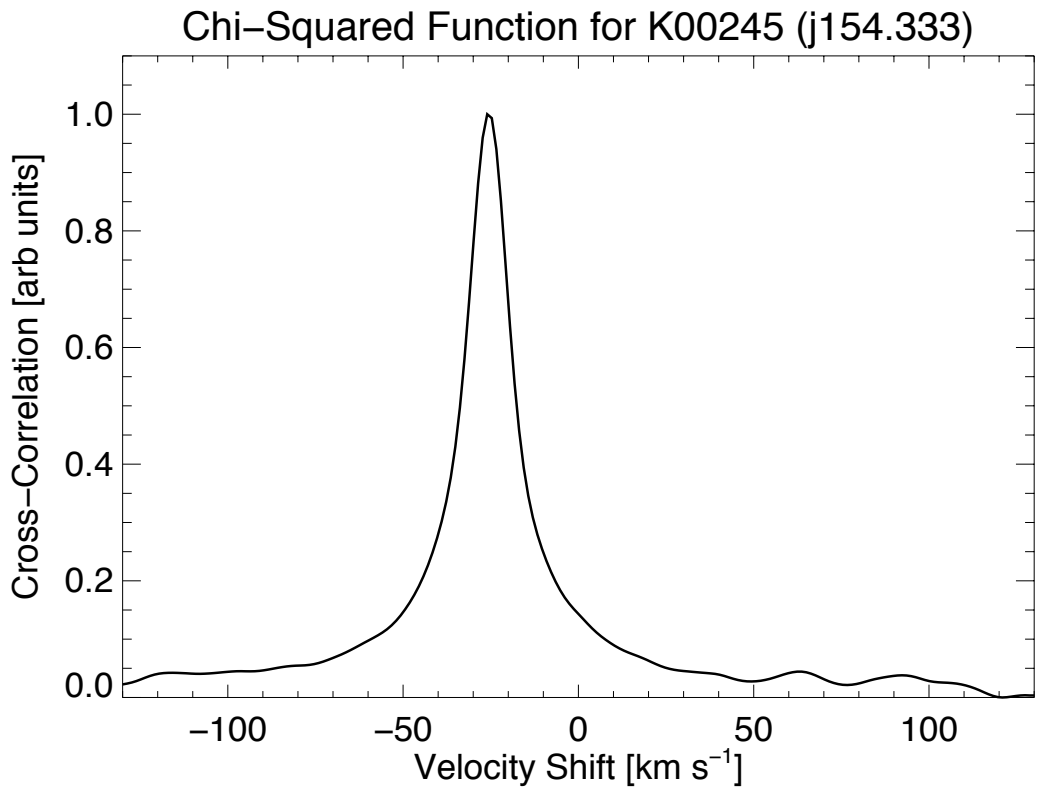

**Figure S3 Cross-correlation function of 25 spectra with a solar template.** High-resolution spectra were obtained using the HIRES spectrograph on the Keck Telescope. In order to determine whether Kepler-37 is blended with another source we cross-correlated these spectra with a solar template. A source with a different radial velocity from Kepler-37 would have appeared as a peak in the plot away from the main peak. The only peaks seen are symmetric about -30.1 km s$^{-1}$ velocity (the average velocity of the star), indicating these are caused by coincidental over-lapping of lines in the spectra. The lack of suspicious peaks above the noise rules out background stars brighter than ΔKp = 4.0 with respect to Kepler-37 with a different absolute radial velocity.

**4. High contrast imaging**

We obtained several sets of high contrast images using a variety of techniques. Adaptive optics (AO) images of Kepler-37 were taken using the ARIES camera on the MMT[43] in Ks and J bands using the F/30 mode on 2010 May 03 and again using the PHARO instrument on the Palomar 200-inch telescope on 2012 June 06 with a Br-γ filter. No source is seen in these images down to a depth of 7.1 mags at separations of 0.5 arcsec and further and 9.3 mags at 1.0 arcsec and further where quoted magnitudes are at 3-σ upper limits.

Kepler-37 was also observed on two separate occasions with the Differential Speckle Survey Instrument[44,45] (DSSI). Firstly, at the WIYN 3.5-m telescope located on KittPeak on 20 June 2010 in V (562nm) and R (692 nm) bandpasses44. Our speckle observing setup at WIYN consisted of 3 sets of 1000 40 msec exposures co-added together. Kepler-37 was seen as a single source to a depth of 3.98 mags and 4.73 mags (value quoted for radius = 0.2 arcsec) in V and R respectively. Kepler-37 was again observed with DSSI on 27 July 2012 UT at the Gemini-N 8-m telescope[46] using R (692 nm) and I (880 nm) filters. One set of 1000 60 msec exposures was obtained. The target star was again seen as a single object to depths of 5.04 mags in R and 5.01 mags in I (again at a radius of 0.2 arcsec). Magnitude limits quoted for speckle imaging are 5-σ upper limits.

Finally, Kepler-37 was observed using the Lucky Imaging technique at the Calar Alto Observatory 2.2m telescope and the Astralux instrument under good seeing conditions (0.8 arcsec) on 2012 May 27, using a Sloan *i* filter. This configuration allowed us to perform diffraction limited imaging of Kepler-37. We used the full camera array (24×24 arcsec) to cover the whole Kepler PSF. We took 30000 frames of 30 milliseconds of exposure time (well below the typical timescale on which atmospheric turbulence changes). We used the AstraLux pipeline to reduce and combine the image frames, which estimates the quality of each individual science frame in order to select the 1% frames of the with the highest Strehl ratios and performs the stacking, producing final images with a pixel scale of 0.02327 arcsec/pixel.

According to our sensitivity measurements on the final image, we did not detect any source at the 3-σ level for objects two magnitudes fainter at 0.25 arcsec from the target star, 4 magnitudes at 0.4 arcsec, 6 magnitudes at 0.5 arcsec, and 7 magnitudes at angular separations greater than 1.4 arcsec. No objects within 8 arcsec are found for Kepler-37 within these sensitivity limits.

**5. Transits of Kepler-37d at 4.5 microns with Warm-Spitzer**
Kepler-37d was observed during one transit with Warm Spitzer/IRAC[47,48] at 4.5 μm. These observations occurred on 2010 November 25 and the visit lasted 11.4 h. The data were gathered in subarray mode (32 x 32 pixels) with an exposure time of 2 s per image, which yielded 22400 images. We produced the photometric time series[49] by finding the centroid position of the stellar point spread function (PSF) and performing aperture photometry using a circular aperture on individual exposures. The images used are the Basic Calibrated Data (BCD) delivered by the Spitzer archive. These files are corrected for dark current, flat-fielding, detector non-linearity and converted into flux units. We converted the pixel intensities to electrons using the information given in the detector gain and exposure time provided in the FITS headers to facilitates the evaluation of the photometric errors. We then converted to UTC-based BJD[50] and corrected for transient pixels in each individual image using a 20-point sliding median filter of the pixel intensity versus time. For this step, we compared each pixel's intensity to the median of the 10 preceding and 10 following exposures at the same pixel position and we replaced outliers greater than 4-σ with its median value. The centroid position of the stellar PSF is determined and then we perform an aperture photometry extraction with a circular aperture of variable radius, using radii of 1.5 to 8 pixels, in 0.5 steps. The propagated uncertainties are derived as a function of the aperture radius; we adopt the one that provides the smallest errors. We find that the transit depths and errors vary only weakly with the aperture radius. The optimal aperture is found to be at 3.5 pixels. We estimated the background by fitting a Gaussian to the central region of the histogram of counts from the full subarray. The centre of the Gaussian fit was adopted as the residual background intensity. The contribution of the background to the total flux from the star is low, from 0.1% to 0.6% depending of the images. Therefore, photometric errors are not dominated by fluctuations in the background. We used a sliding median filter to select and trim outliers in flux and positions greater than 5σ. We also discarded the first half-hour of observations, which are affected by a significant telescope jitter before stabilization. The final number of photometric measurements used is 19780.

The raw time series are presented in the top panels of Figure 5. We find a typical signal-to-noise ratio (S/N) of 365 per image which corresponds to 90% of the theoretical signal-to-noise. Therefore, the noise is dominated by Poisson photon noise. We used a transit light curve model multiplied by instrumental decorrelation functions to measure the transit parameters and their uncertainties[51] and compute a transit light curve[17]. For these Spitzer observations, the transit model depends on one parameter: the planet-to-star radius ratio. The other transit parameters are set fixed to the value derived from the Kepler-37d lightcurve. The limb-darkening coefficients are set to zero since these Spitzer lightcurves do not have enough photometric precision. The Spitzer/IRAC photometry is known to be systematically affected by the so-called *pixel-phase effect*[52,53]. We decorrelated our signal in each channel using a linear function of time for the baseline (two parameters) and a quadratic function of the PSF position (four parameters) to correct the data for each channel. We checked that adding parameters to the quadratic function of the PSF position does not improve the fit significantly. We performed a simultaneous Levenberg-Marquardt least-squares fit[54] to the data to determine the transit and instrumental model parameters (7 in total). The errors on each photometric point were assumed to be identical, and were set to the *rms* of the residuals of the initial best-fit obtained. To obtain an estimate of the correlated and systematic errors[55] in our measurements, we use the residual permutation bootstrap, or ``Prayer Bead'', method[56]. In this method, the residuals of the initial fit are shifted systematically and sequentially by one frame, and then added to the transit light curve model before fitting again. We allow asymmetric error bars spanning 34% of the points above and below the median of the distributions to derive the 1-σ uncertainties for each parameters[57].

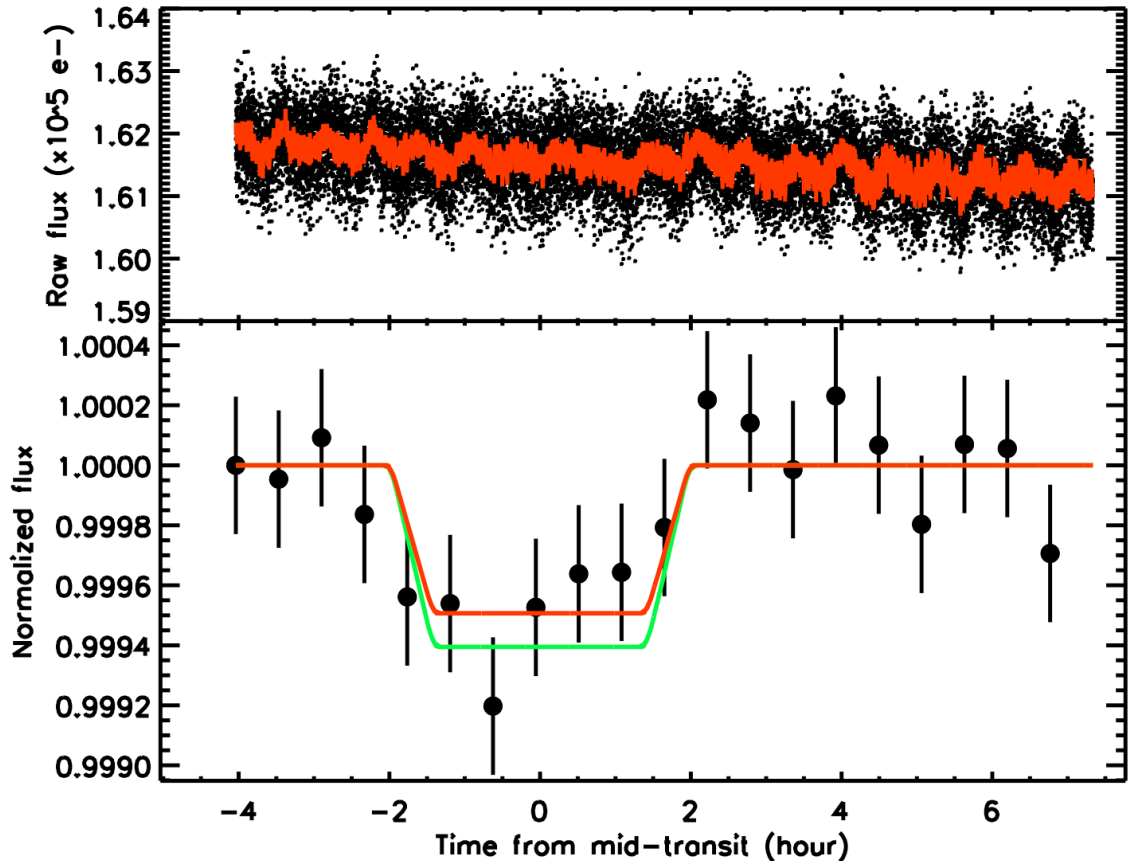

**Figure S4 Spitzer transit light-curve of Kepler-37d observed in the IRAC bandpass at 4.5 μm.** Top panel: raw (unbinned) transit light-curve. The red solid line corresponds to the best-fitting model which includes the time, position instrumental decorrelations as well as the model for the planetary transit. Bottom panel: corrected, normalized and binned by 30 minutes transit light-curve with the transit best-fit plotted in red and the transit shape expected from the Kepler observations overplotted as a green line. The two models agree at close to a 1-sigma level.}

We measured the transit depth at 4.5 μm of 510±60 ppm for Kepler-37d. This value measured in the Spitzer bandpass is in agreement at around the 1-σ level compared to the transit depth measured in the Kepler bandpass ($574.9^{+3.2}_{-3.5}$ ppm). With no evidence for a transit depth that depends on wavelength, the available data is consistent with a dark planetary object.

## 6. Constraints on the composition of Kepler-37b from mass loss

Given Kepler-37b's extraordinarily small size, its mass is likely to be low. Combined with its highly irradiated surface due to the short orbital period, this means that any volatiles are likely going to be extremely vulnerable to atmospheric escape due to heating by XUV (1-1200 Å) photons. In order to examine whether Kepler-37b could have a volatile atmosphere, we used coupled thermal evolution and energy limited mass loss models[57].

If Kepler-37b currently has 50% of its mass atop an Earth-like core, then the planet would be 6.6 x $10^{-4}$ $M_{\oplus}$. If we assume that the star is currently 6 Gyr old and a standard mass loss efficiency of 10%[58,59], then Kepler-37b would lose this entire water layer within the next 5 Myr. If the star is younger, then this timescale becomes even shorter[60,61]. Likewise, if we assume that Kepler-37b has an Earth-like composition today but was initially 90% water when it finished forming at ~10 Myr[62], then it would have lost its entire water envelope by the time it was 30 Myr old. Any H/He envelope around Kepler-37b would be even more vulnerable to mass loss than water.

Even if we imagine a planet that evolved inwards toward the star later in the planetary formation avoiding any early time high UV flux, a planet with an initial composition of 50% water would lose the entire envelope in 20Myr. It is therefore very likely that the surface of Kepler-37b is rocky.

**7. Other planet candidates in the photometric data of Kepler-37**
The most recent Kepler planet catalogue[11] lists four planet candidates orbiting Kepler-37: KOI-245.01, .02, .03 and .04. KOI-245.01, .02, .03 refer to Kepler-37d, c and b, respectively. We do not trust that KOI-245.04 is a valid planet candidate because incorporating more data obtained since the release of the most recent planet candidate catalogue has resulted in a decrease in the signal to noise of the folded transit signal. This is a strong indication that the signal originally interpreted as being a transiting planet is likely to be caused by either random noise, or correlated noise either from the star (i.e. starspot activity) or instrumental artefacts such as sudden pixel sensitivity dropouts[63].

## 8. Stellar and planetary parameters for the Kepler-37 system

In the table below we show the measured and derived star and planet parameters of Kepler-37. The stellar parameters come from the spectroscopic and asteroseismic analysis while the planet parameters mainly come from our transit model with the exception of transit depth measured using Spitzer data.

**Table S1 Parameters of the star and planets in the Kepler-37 system.**

| Stellar Parameters | |
|---|---|
| Brightness (Kp) | 9.701 |
| Effective temperature, $T_{eff}$ (K) | 5417 ± 75 |
| Surface gravity, log g (dex, cgs units) | 4.5667 ± 0.0065 |
| Metallicity, [m/H] | -0.32 ± 0.07 |
| Average large frequency spacing, $\langle \Delta \nu \rangle$ (μHz) | 178.7 ± 1.4 |
| Average small frequency spacing, $\langle \delta \nu_{01} \rangle$ (μHz) | 4.08 ± 0.17 |
| Mean density, $\langle \rho \rangle$ (g cm$^{-3}$) | 2.458 ± 0.046 |
| Absolute radial velocity (km s$^{-1}$) | -30.1 ± 0.2 |
| Project rotational velocity, $V \sin i$ (km s$^{-1}$) | 1.1 ± 1.1 |
| Age (Gyr) | ~6 |
| Distance (pc) | ~66 |
| Mass ($M_\odot$) | 0.803 ± 0.068 |
| Radius ($R_\odot$) | 0.770 ± 0.026 |
| Limb-darkening coefficients, {a1, a2, a3, a4} | {0.3944, 0.2971, 0.270, -0.2203} |

| Planet parameters | Kepler-37b | Kepler-37c | Kepler-37d |
|---|---|---|---|
| Orbital period (days) | $13.367308^{+0.000058}_{-0.000085}$ | $21.301886^{+0.000046}_{-0.000044}$ | $39.792187^{+0.000040}_{-0.000043}$ |
| Planet-to-star radius ratio | $0.00360^{+0.00058}_{-0.00085}$ | $0.00877^{+0.00037}_{-0.00061}$ | $0.02359^{+0.00025}_{-0.00043}$ |
| Transit epoch (BJD – 2454833) | $184.03271^{+0.00034}_{-0.00089}$ | $191.83816^{+0.00036}_{-0.00078}$ | $175.24979^{+0.00043}_{-0.00049}$ |
| $e \cos \omega$ | $-0.47^{+0.63}_{-0.31}$ | $-0.04^{+0.12}_{-0.15}$ | $0.03^{+0.12}_{-0.12}$ |
| $e \sin \omega$ | $-0.54^{+0.16}_{-0.33}$ | $-0.09^{+0.12}_{-0.10}$ | $-0.098^{+0.033}_{-0.025}$ |
| Impact parameter | $0.71^{+0.23}_{-0.19}$ | $0.66^{+0.21}_{-0.15}$ | $0.715^{+0.032}_{-0.048}$ |
| Inclination (°) | $88.63^{+0.30}_{-0.53}$ | $89.07^{+0.19}_{-0.33}$ | $89.335^{+0.043}_{-0.047}$ |
| Scaled semi-major axis | $28.526^{+0.088}_{-0.096}$ | $38.92^{+0.12}_{-0.13}$ | $59.03^{+0.18}_{-0.20}$ |
| Semi-major axis (au) | $0.1003^{+0.0008}_{-0.0011}$ | $0.1368^{+0.0011}_{-0.0014}$ | $0.2076^{+0.0016}_{-0.0022}$ |
| Transit depth (ppm) | $11.9^{+2.6}_{-3.1}$ | $81.1^{+2.6}_{-2.9}$ | $574.9^{+3.2}_{-3.5}$ |
| Radius ($R_\oplus$) | $0.303^{+0.053}_{-0.073}$ | $0.742^{+0.065}_{-0.083}$ | $1.99^{+0.11}_{-0.14}$ |

**References**
31. Chaplin, W. J. *et al.* Ensemble asteroseismology of solar-type stars with the NASA Kepler Mission. *Science*, **332**, 213 – 216 (2011)
32. Verner, G. A. *et al.* Global asteroseismic properties of solar-like oscillations observed by Kepler: a comparison of complementary analysis methods. *Mon. Not. R. Astron. Soc.* **415**, 3539–3551 (2011)
33. Vogt,S.S *et al.* HIRES: the high-resolution echelle spectrometer on the Keck 10-m Telescope. *Proc. SPIE* **2198**, 362 (1994)
34. Stello, D. *et al.* Radius determination of solar-type stars using asteroseismology: What to expect from the Kepler Mission. *Astrophys. J.* **700**, 1589–1602 (2009)
35. Gai, N. *et al.* An in-depth study of grid-based asteroseismic analysis, *Astrophys. J.* **730**, 63 (2011)
36. Quirion, P.-O. *et al.* Automatic determination of stellar parameters via asteroseismology of stochastically oscillating stars: comparison with direct measurements. *Astrophys. J.* **725**, 2176–2189 (2010)
37. Torres, G., *et al.* Improved Spectroscopic Parameters for Transiting Planet Hosts. *Astrophys. J.* **757**, 161
38. Bryson, S. T. *et al.* The Kepler Pixel Response Function. *Astrophys. J.* **713**, 97–102 (2010)
39. Howell, S. B. *et al.* Kepler-21b: A 1.6REarth Planet Transiting the Bright Oscillating F Sub- giant Star HD 179070. *Astrophys. J.* **746**, 123 (2012)
40. Lomb, N. R. Least-squares frequency analysis of unequally spaced data. *Astrophys. Space Sci.* **39**, 447-462 (1976)
41. Scargle, J. D. Studies in astronomical time series analysis. II - Statistical aspects of spectral analysis of unevenly spaced data. *Astrophys. J.* **263**, 835-853 (1982)
42. Isaacson, H. & Fischer, D. Chromospheric Activity and Jitter Measurements for 2630 Stars on the California Planet Search. *Astrophys. J.* **725**, 875-885 (2010)
43. Adams, E. R. *et al.* Adaptive Optics Images of Ke- pler Objects of Interest. *Astron. J.* **144**, 42 (2012)
44. Howell, S. B., Everett, M. E., Sherry, W., Horch, E. & Ciardi, D. R. Speckle Camera Observations for the NASA Kepler Mission Follow-up Program. *Astron. J.* **142**, 19 (2011)
45. Horch, E. P. *et al.* CCD Speckle Observations of Binary Stars with the WIYN Telescope. VI. Measures During 2007-2008. *Astron. J.* **139**, 205–215 (2010)
46. Horch, E., Howell, S. B. Everett, M. E. &Ciardi, D. R. Observations of Binary Stars with the DSSI IV. Observations of Kepler, CoRoT, and Hipparcos Stars from Gemini North. *Astron. J.* **144**, 165 (2012)

47. Werner, M. W. *et al.* The Spitzer Space Telescope Mission. *Astrophys. J. Suppl.* **154**, 1–9 (2004)
48. Fazio, G. G. *et al.* The Infrared Array Camera (IRAC) for the Spitzer Space Telescope. *Astrophys. J. Suppl.* **154**, 10 (2004)
49. Désert, J.-M., Lecavelier des Etangs, A., Hébrard, G., Sing, D. K., Ehrenreich, D., Ferlet, R. & Vidal-Madjar, A. Search for Carbon Monoxide in the Atmosphere of the Transiting Exoplanet HD 189733b. *Astrophys. J.* **699**, 478 (2009)
50. Eastman, J., Siverd, R., & Gaudi, B. S. Achieving Better Than 1 Minute Accuracy in the Heliocentric and Barycentric Julian Dates. *Publ. Astron. Soc. Pac.* **122**, 935–946 (2010)
51. Désert, J.-M. *et al.* Transit spectrophotometry of the exoplanet HD 189733b. II. New Spitzer observations at 3.6 μm, *Astron. Astrophys.* **526**, A12 (2011)
52. Charbonneau, D. *et al.* Detection of Thermal Emission from an Extrasolar Planet. *Astrophys. J.* **626**, 523 (2005)
53. Knutson, H. A., Charbonneau, D., Allen, L. E., Burrows, A., & Megeath, S. T. The 3.6-8.0 μm Broadband Emission Spectrum of HD 209458b: Evidence for an Atmospheric Temperature Inversion. *Astrophys. J.* **673**, 526–531 (2008)
54. Markwardt, C.B. Non-linear Least-squares Fitting in IDL with MPFIT. *Astronomical Society of the Pacific Conference Series* **411**, 251 (2009)
55. Pont, F., Zucker, S., & Queloz, D. The effect of red noise on planetary transit detection. *Mon. Not. R. Astron. Soc.* **373**, 231–242 (2006)
56. Désert, J.-M. *et al.* The Atmospheres of the Hot-Jupiters Kepler-5b and Kepler-6b Observed during Occultations with Warm-Spitzer and Kepler. *Astrophys. J.* **197**, 11 (2011)
57. Lopez, E. D., Fortney, J. J., & Miller, N. How Thermal Evolution and Mass Loss Sculpt Populations of Super-Earths and Sub-Neptunes: Application to the Kepler-11 System and Beyond. *Astrophys. J.* **761**, L59 (2012)
58. Erkaev, N. V., Kulikov, Y. N., Lammer, H., Selsis, F., Langmayr, D., Jaritz, G. F. & Biernat, H. K. Roche lobe effects on the atmospheric loss from "Hot Jupiters". *Astron. Astrophys.* **472**, 329–334 (2007)
59. Murray-Clay, R. A., Chiang, E. I. & Murray, N. Atmospheric Escape From Hot Jupiters. *Astrophys. J.* **693**, 23–42 (2009)
60. Ribas, I., Guinan, E. F., Güdel, M. & Audard, M. Evolution of the Solar Activity over Time and Effects on Planetary Atmospheres. I. High-Energy Irradiances (1-1700 Å ). *Astrophys. J.* **622** 680–694 (2005)
61. Sanz-Forcada, J. *et al.* Estimation of the XUV radiation onto close planets and their evaporation. *Astron. Astrophys.* **532**, A6 (2011)
62. Calvet, N. *et al.* Evidence for a Developing Gap in a 10 Myr Old Protoplanetary Disk. *Astrophys. J.* **568**, 1008–1016 (2002)

63. Smith, J. C. *et al.* Kepler Presearch Data Conditioning II - A Bayesian Approach to Systematic Error Correction. *Publ. Astron. Soc. Pac.* **124**, 1000 (2012)